%% file: 00main.tex
\documentclass[conference,a4paper]{IEEEtran}
\IEEEoverridecommandlockouts
\usepackage{cite}
\usepackage{amsmath,amssymb,amsfonts}
\usepackage{algorithmic}
\usepackage{graphicx}
\usepackage{textcomp}
\usepackage{subcaption}
\usepackage{booktabs}
\usepackage{multirow}
\usepackage{tikz}
\usetikzlibrary{positioning}

\newcommand{\myvec}[1]{\mathbf{#1}}

\usepackage{etoolbox}
\usepackage{xcolor}

\newcommand\copyrighttext{%
  \footnotesize \textcopyright 2020 IEEE. Personal use of this material is permitted.
  Permission from IEEE must be obtained for all other uses, in any current or future 
  media, including reprinting/republishing this material for advertising or promotional 
  purposes, creating new collective works, for resale or redistribution to servers or 
  lists, or reuse of any copyrighted component of this work in other works. This paper has been accepted for publication in the 2020 Twelfth International Conference on Quality of Multimedia Experience (QoMEX).
	}
\newcommand\copyrightnotice{%
\begin{tikzpicture}[remember picture,overlay]
\node[anchor=south,yshift=10pt] at (current page.south) {\fbox{\parbox{\dimexpr\textwidth-\fboxsep-\fboxrule\relax}{\copyrighttext}}};
\end{tikzpicture}%
}

\newcommand{\tmpcomment}[3]{{\emph{\textcolor{#1}{#2: #3}}}}

\newcommand{\babak}[1]{\tmpcomment{purple}{Babak}{#1}}

\newcommand{\rafael}[1]{\tmpcomment{green}{Rafael}{#1}}
\newcommand{\tobias}[1]{\tmpcomment{brown}{Tobias}{#1}}
\newcommand{\matthias}[1]{\tmpcomment{blue}{Matthias}{#1}}

\newtoggle{hidecomments}
\toggletrue{hidecomments}
\iftoggle{hidecomments}{
	\renewcommand{\tmpcomment}[3]{}
}{
}
\usepackage{url} 

\newcommand{\accessedDate}{March~2020}
\newcommand{\urlfootnote}[1]{\footnote{\url{#1} Accessed \accessedDate}}
\input{custom}

\def\BibTeX{{\rm B\kern-.05em{\sc i\kern-.025em b}\kern-.08em
    T\kern-.1667em\lower.7ex\hbox{E}\kern-.125emX}}
\begin{document}

\IEEEoverridecommandlockouts
%\IEEEpubid{\makebox[\columnwidth]{
%\scriptsize{978-1-7281-5965-2/20/\$31.00 \copyright2020 IEEE} 
%}\hspace{\columnsep}\makebox[\columnwidth]{}}

\title{Impact of the Number of Votes on the Reliability and Validity of Subjective Speech Quality Assessment in the Crowdsourcing Approach}

\author{
    \IEEEauthorblockN{Babak Naderi\IEEEauthorrefmark{1}, Tobias Ho{\ss}feld\IEEEauthorrefmark{2}, Matthias Hirth\IEEEauthorrefmark{3}, Florian Metzger\IEEEauthorrefmark{2}, Sebastian M{\"o}ller\IEEEauthorrefmark{1}\IEEEauthorrefmark{4},
    Rafael Zequeira Jim{\'e}nez\IEEEauthorrefmark{1}}
    \IEEEauthorblockA{\IEEEauthorrefmark{1}Quality and Usability Lab, Technische Universit{\"a}t Berlin, Germany, forename.surname@tu-berlin.de}
    \IEEEauthorblockA{\IEEEauthorrefmark{2}University of Würzburg, Germany, forename.surname@uni-wuerzburg.de}
    \IEEEauthorblockA{\IEEEauthorrefmark{3}Technische Universit{\"a}t Ilmenau, Germany, matthias.hirth@tu-ilmenau.de}
    \IEEEauthorblockA{\IEEEauthorrefmark{4}DFKI Projektb{\"u}ro Berlin, Germany, sebastian.moeller@dfki.de}
    
}  

\pagestyle{plain}

\maketitle
\copyrightnotice

\begin{abstract}
The subjective quality of transmitted speech is traditionally assessed in a controlled laboratory environment according to ITU-T Rec. P.800. 
In turn, with crowdsourcing, crowdworkers participate in a subjective online experiment using their own listening device, and in their own working environment.
	Despite such less controllable conditions, the increased use of crowdsourcing micro-task platforms for quality assessment tasks has pushed a high demand for standardized methods, resulting in ITU-T Rec. P.808. This work investigates the impact of the number of judgments on the reliability and the validity of quality ratings collected through crowdsourcing-based speech quality assessments, as an input to ITU-T Rec. P.808 \matthias{We might remove this (input to ITU...), as this is not relevant for the scientific contribution of this paper. IMHO}.
Three crowdsourcing experiments on different platforms were conducted to evaluate the overall quality of three different speech datasets, using the Absolute Category Rating procedure. 
For each dataset, the Mean Opinion Scores~(MOS) are calculated using differing numbers of crowdsourcing judgements.
Then the results are compared to MOS values collected in a standard laboratory experiment, to assess the validity of crowdsourcing approach as a function of number of votes. In addition, the reliability of the average scores is analyzed by checking inter-rater reliability, gain in certainty, and the confidence of the MOS.
The results provide a suggestion on the required number of votes per condition, and allow to model its impact on validity and reliability. 

%\todo{Alternative title: Recommended number of votes for crowdsourced subjective speech quality assessment in ITU-T Rec. P.808}
%\matthias{I would prefer a title without mentioning the recommendation. Mentioning it might make the paper less attractive to people outside the ITU-T community.} 

%\rafael{I also prefer it without P.808, this is my proposal:}
%\rafael{Alternative title: "Impact of the Number of Votes on the Reliability and Validity of Subjective Speech Quality Assessment in Crowdsourcing"}

\end{abstract}

\begin{IEEEkeywords}
crowdsourcing, speech, quality assessment, reliability, validity
\end{IEEEkeywords}
%\begin{tikzpicture}[overlay, remember picture]
%\path (current page.north) node (anchor) {};
%\node [below=of anchor] {%
%2020 Twelfth International Conference on Quality of Multimedia Experience (QoMEX)};
%\end{tikzpicture}
%\babak{check for: (lab and laboratory), (ratings and votes), (CS, crowdsourcing)}

\input{01intro.tex}
\input{02method.tex}
\input{03results.tex}

\input{04discussion.tex}

\bibliographystyle{IEEEtran}
\bibliography{library}

\end{document}

%% file: custom.tex
\usepackage{xcolor}

%% file: 01intro.tex
\section{Introduction}

The quality of speech transmitted through communication networks is commonly seen as a result of a perception and judgment process~\cite{Jekosch05}.
During the assessment, a listener perceives an acoustic event, and analyzes it according to a number of criteria. 
These criteria are partially defined externally (e.g. by the test task), and partially by the internal reference which is built through prior exposure to this and other acoustic events.
Frequently, the task of the assessment requires some quantitative scaling of the entity, e.g. a number related to the overall quality, intelligibility, coloration, or alike, usually a rating on a category scale.
As a result of individual differences in the perception process and the internal reference, the same acoustic event may result in different quantitative judgments.
In addition, the acoustic event may differ because of differences in the experimental setup for each listener, or across repetitions of the same experiment.

The uncertainty of the quality judgments --- and thus of the perceived quality --- is inherent to every measurement process, be it based on human subjects or on technical instruments.
The goodness of such measurements is commonly expressed by two main criteria: The validity, i.e. the method should be able to measure what it is intended to measure, and the reliability, i.e. the method should be able to provide stable results across repeats of the same measurement~\cite{Lienert89}.
Other goodness criteria are e.g. the sensitivity (able to measure small variations in what it is intended to measure), its objectivity (reach inter-individual agreement on the measurement results), its robustness (able to provide results independent from variables that are extraneous to the construct being measured), as well as its efficiency (provide good results with limited efforts invested)~\cite{Lienert89}.

When it comes to speech quality assessments, most recommended methods make use of subjective listening experiments carried out under controlled listening conditions.
These conditions include the selection of test participants, the acoustic characteristics of the test environment, the processing of the speech stimuli triggering the acoustic event, the listening devices used for presenting these events, the test task and test design, the rating scale(s) and procedures.
These recommendations are summarized by the Telecommunication Standardization Sector of the International Telecommunication Union (ITU-T) in Rec. P.800~\cite{ITU-P800}. The majority of speech quality assessments in the literature follow this recommendation.

\babak{Despite its common usage, the recommendations given in ITU-T Rec. P.800 have been disputed.
One point of criticism is the calculation of an arithmetic mean value of ratings which have been obtained on a scale which possesses, in most cases, not more than a rank order property.
For such a scale, the distances between scale labels (categories) are not properly defined, and they are not equidistant.
A second point of criticism is the rather artificial test situation:}
The laboratory (lab) tests are criticized because of their artificial test situation: Whereas average users of telecommunication services listen to speech in acoustically spoiled conditions, with rather sub-optimal listening devices, the lab test situation is meant to increase the sensitivity and robustness of the measurement process.
As a consequence, the ecological validity of the measurement suffers, as the assessment situation does not reflect ``normal'' service usage any more.
In addition, conducting lab tests is expensive and time intensive.

With the rise of crowdsourcing (CS) micro-task platforms, proposals have been made to transfer listening quality assessments and other assessments of quality to the crowd~\cite{naderi2015effect, polzehl2015robustness, Egger-Lampl2017a, ZequeiraJimenez2018}.
The use of online portals might attract a larger and more diverse group of listeners, thus better representing service users -- this might increase validity.
In addition, a larger number of usage situations, including listening devices, background noises, distractions from the listening process, etc., might increase the realism of the measurement, and thus ecological validity.
%\rafael{"And, foremost, online measurements are quicker and easier to set up, especially when crowdsourcing platforms facilitate the recruiting of listeners."} %\rafael{proposed to be deleted if more space is needed}
In turn, the lack of control puts severe doubts on the reliability of the measurement.
These and other characteristics are summarized in ITU-T Rec. P.808~\cite{ITU-P808}, which has been set up with the aim of increasing the reliability and acceptability of crowdsourcing-based speech quality measurements.

Whereas there are several established methods for analyzing the reliability of speech quality measurements, their validity is more difficult to estimate.
One could take the position that valid quality judgments can only be obtained from a fully representative choice of users, user devices, and usage situations.
Following this argument, a crowdsourcing-based assessment would be more ecologically valid than a lab-based assessment.
Unfortunately, there are no reliable reports on the diversity of users, devices and listening situations which would justify that a standard crowdsourcing-based study would reach that aim.
On the other hand, lab tests are currently the most used method for speech quality assessment, thus the result of such an assessment could be seen as a type of ``gold standard''.

%It is the aim of the present paper to assess some characteristics of speech quality assessments in the lab and in the crowd.
It is the aim of the present paper to assess the impact of the number of judgments on the reliability and validity of overall quality judgments in the crowd.
In doing so, we will examine the validity, with lab tests as a ``gold standard'', and calculate the rank-order correlation, and an absolute difference as a criterion.
Reliability will be checked in terms of inter-rater reliability, gain in certainty, and confidence of the MOS.

\rafael{
The paper is structured as follows:
Section~\ref{sec:method} will describe the methods used for our analysis, including the laboratory and crowdsourcing experiments, and simulations.
Section~\ref{sec:results} will report the results and analyze them regarding validity and reliability.
Section~\ref{sec:discussion} will conclude with a general discussion and implications for ITU-T Rec. P.808.
}
 \rafael{Rafael: proposed to be deleted if more space is needed}

%% file: 02method.tex
\section{Methodology}
\label{sec:method}

%In the following, the speech datasets and the three crowdsourcing studies that were conducted according to the ITU-T Rec. P.808~\cite{ITU-P808} are described.
%Afterwards, we explain our simulation method.

%\subsection{Speech Datasets}
Three datasets from the dataset pool of the ITU-T Rec. P.863~\cite{ITU-P863} competition were selected, namely 401, 501 and 701.
They were prepared according to ITU-T Rec. P.800. Access to them was kindly provided to us for these evaluations.
The datasets include two different languages and study designs and cover a variety of degradations.
For each dataset, the results of lab-based experiments were also provided by corresponding contributors.
Tab.~\ref{tab:datasets} summarizes the source materials and the design of the lab assessments. 

\input{tabs/tab_datasets.tex}

\subsection{Crowdsourcing Experiments}
One crowdsourcing test was conducted for each of the datasets.
The tests followed the ITU-T Rec. P.808, and consisted of a qualification, a training, and a rating part.
Additionally, different data cleansing methods were applied based on the instructions given in ITU-T Rec. P.808. 
In the following, we briefly summarise the individual tests\footnote{We made the crowdsourcing ratings openly available at https://github.com/hossfeld/crowdsourced-speech-quality}.

\newcommand{\csmt}{\textit{CS~401}}
\newcommand{\cscw}{\textit{CS~501}}
\newcommand{\csmw}{\textit{CS~701}}

\subsubsection{CS 401}
\csmt\ was performed on Amazon Mechanical Turk\urlfootnote{https://www.mturk.com} using its internal template engine.
Only US workers with a task approval rate above 98\% and more than $500$ accepted jobs were allowed to participate.
%\rafael{
The qualification tasks consisted of a self assessment of the workers' hearing capabilities and a modified version of digit-triplet test \cite{smits2004development}
as a hearing test.
% Rafael: "self assessment of the workers' hearing capabilities" and "version of digit-triplet test" is the same thing right?
% Babak: no, they are not
Here, the workers had to listen to five stimuli with a signal-to-noise ratio (SNR) of -11.2~dB and type in the numbers they heard. 
This SNR was chosen to reach high true positive rate.
% while previous study suggested to use threshold of -9.3~dB SNR for German, -11.2~dB SNR for Dutch, and -10.5~dB SNR for French digit-triplet test to find normal hearing participants \cite{buschermohle2015german}.
%}
% Rafael: I propose to omit all the things related to the digit triplet test, since those results are not further analyzed. Better just include a simpler sentence about the qualification step skipping the info about the digit triplet test
From $227$ participants, $187$ workers successfully passed the qualification. Then, $100$ were randomly selected for the training and rating task.
In the training task, the workers could familiarize themselves with the task interface and with different stimuli representing the entire quality range.
The rating task consisted of $10$ stimuli, one trapping stimulus, a stereo test and environment suitability test. %to check if both earpods were used by participants. 
We aimed at collecting $10$ votes per stimulus.
Overall, we collected $1160$ rating tasks, each including $10$ votes, from $71$ unique workers.
% @Rafael: In total $1042$ rating tasks from $68$ workers were accepted for further analysis, after removing responses with invalid trapping questions, failed stereo tests, or failed in the environmental screening test.
In total $1042$ rating tasks from $68$ workers were accepted for further analysis, after removing responses with invalid trapping questions and failed stereo or environmental screening test.

\subsubsection{CS 501}
The \cscw\ study was run on Clickworker\urlfootnote{https://www.clickworker.com}, due to its large number of German speaking workers.
Clickworker's templates did not support audio playback at the time of the data collection, thus, an external system was implemented for the tests.
%\footnote{https://gitlab.com/zequeira/NoStimuli-SQA.git}.
% removed for blind review
%\matthias{@Rafael: If you want to publish the code along with this paper, we can upload it to Dropbox and share the link. This enables a double-blinded access for the reviewers, other than GitHub or GitLab.}
% Babak: it was already announced in other papers. We can put a ref to other paper if we get accepted.
% Rafael: we can include the link to the GitLab repository as a footnote.
In this study, the qualification phase was implemented as a German listening comprehension test with three audio stimuli. During the training job, workers had to sum up numbers that were played on the left and right audio channel as a stereo test.  Finally, they listened to five stimuli from the dataset that covered the complete MOS range and thus, serving as anchoring. Workers who successfully answered the math question were allowed to proceed to the rating task.
%
% Rafael (re-writed below): (deleted since these results were not analyzed) "Additionally, workers had to listen to $10$ audio files playing white noise at different frequencies and intensities, and report if they were able to hear something. This allowed for a basic check of the workers' hearing abilities."
%
% \matthias{@Rafael: If we do not use the results from the noise results for further filtering, I suggest to also remove the description.} \rafael{DONE!}
% that was randomly added between the 10th and 15th sample. 
%
The rating job included $15$ stimuli and one trapping question. In total, $5245$ ratings from $64$ unique workers were collected. All workers answered the trapping questions correctly. Then, 136 ratings were identified as extreme outliers (i.e. located at a distance from the median equal or higher than $3.0 \times \text{interquartile range}$) and were removed\babak{~\cite{Hoaglin1987}}.

% \matthias{@Rafael: Is this correct? In the journal it says: beyond an outer fence of boxplot. Further, we might state how you aggregated it, per stimulus?}
% \rafael{the outliers ratings depicted as circles that are beyond an outer fence of boxplot, specifically beyond 2.2 IQR, according to \cite{Hoaglin1987} still might be "good data points", in turn those located at 3.0 IQR or beyond and depicted with asterisks are extreme outliers and can be removed.} 

\subsubsection{CS 701}
The last study, \csmw, was performed on Microworkers\urlfootnote{https://www.microworkers.com}.
Similar to Clickworker, Microworkers' template system also does not support audio playback and an external system was implemented.
% Rafael (re-writed below): The qualification task of \csmw\ did not rely on audio stimuli, in contrast to \csmt\ and \cscw, but is based on self assessment of the workers' hearing capabilities and their surrounding environment.
In contrast to \csmt\ and \cscw, the qualification task of \csmw\ did not rely on audio stimuli, but was based on self-assessment of the workers' hearing capabilities and their surrounding environment.
All workers that completed the qualification task were also allowed to participate in the training and rating task, as a preselection based on the self assessment might encourage other workers to provide incorrect responses to gain access to the succeeding tasks. 
However, only responses from workers that did not report hearing impairments or noisy surroundings were considered.
The training job gave the workers the chance to adjust the playback volume and to familiarize themselves with the task interface.
Further, the workers had to complete a stereo test similar to the one in \cscw. 
After successfully completing the training task, each worker was allowed to submit up to $10$ rating tasks, with each rating task consisting of $10$ stimuli and one trapping stimulus injected at random.
In total $197$ workers completed $1032$ rating jobs.
After removing workers due to self-reported hearing impairments, invalid answers to trapping stimuli and failing the stereo test, $6990$ ratings from 144 workers remained for further analysis.
%\babak{@Matthias: did you use any outlier detection method or worker agreement as a condition to filter participants in case of strict filter ?}
%\matthias{No, I just filtered workers based on hearing ability, noisy environment, failing trap, and failing stereo question. Non of them tasks the acutall ratings into account.}

\begin{figure*}[htbp] 
  \centering
  \begin{subfigure}[b]{0.3\textwidth}
    \centering
    \includegraphics[width=\textwidth]{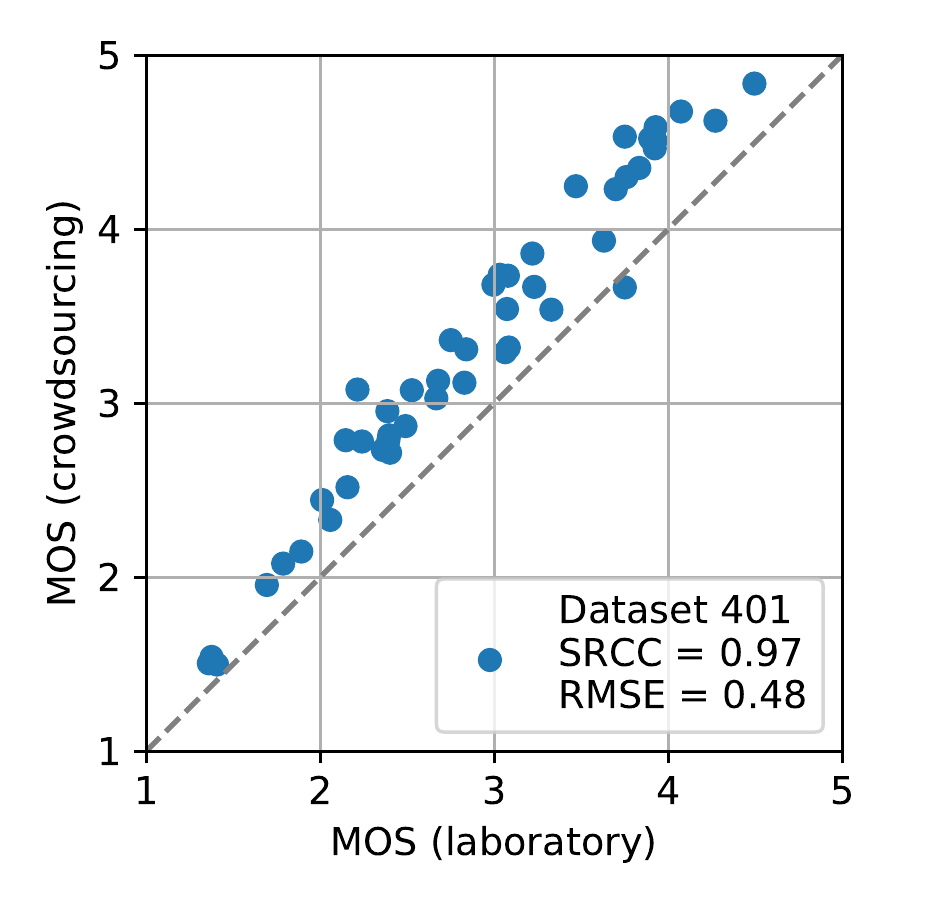} 
    \caption{CS 401} 
    \label{fig7:a} 
  \end{subfigure} 
  ~
  \begin{subfigure}[b]{0.3\textwidth}
    \centering
    \includegraphics[width=\textwidth]{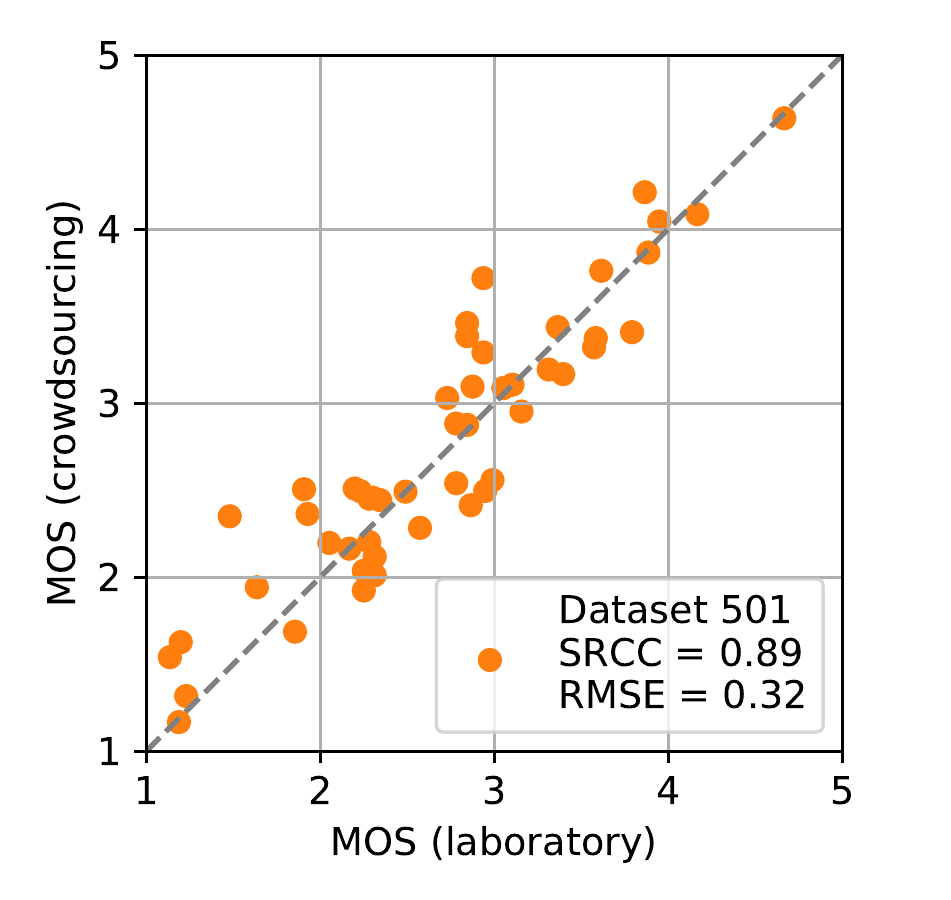}
    \caption{CS 501}
	\label{fig7:b} 
  \end{subfigure} 
  ~
 \begin{subfigure}[b]{0.3\textwidth}
    \centering
    \includegraphics[width=\textwidth]{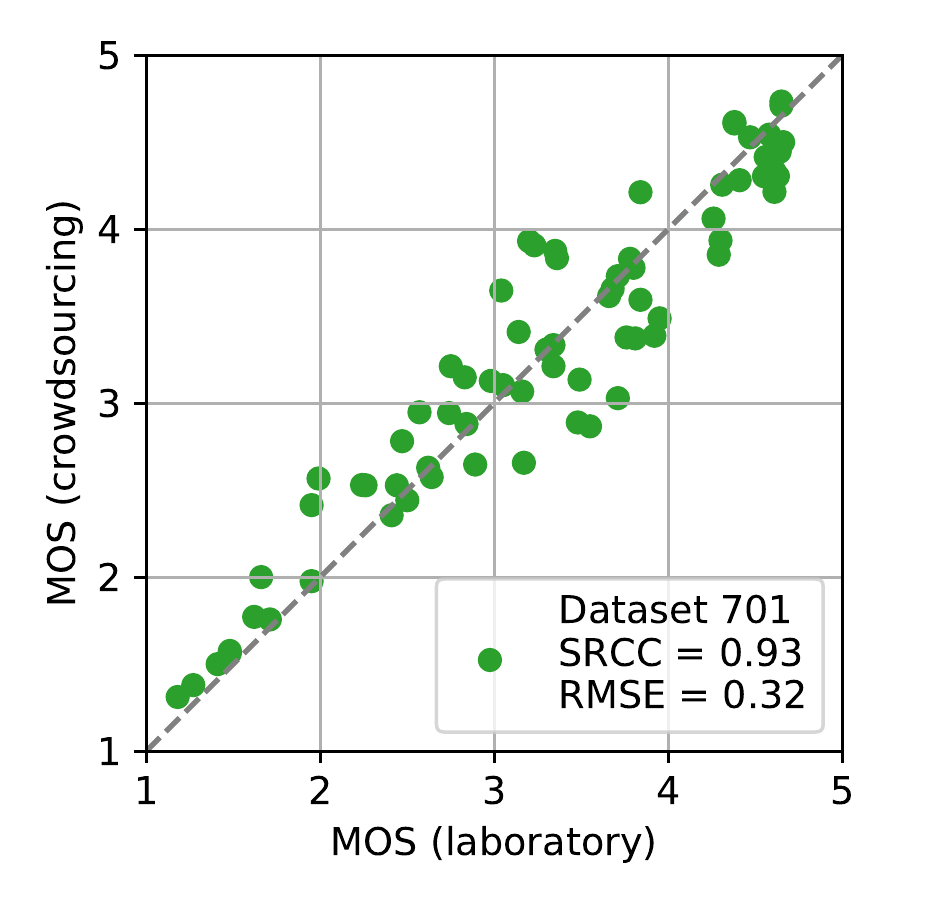}
    \caption{CS 701}
	\label{fig7:c} 
  \end{subfigure} 
  
  \caption{Comparison between MOS values in the laboratory and the crowdsourcing experiments.} %\tobias{I would not use a 1st order mapping. The results are still very good!}}
  \label{fig:result:mapped} 
\end{figure*}

\subsection{Simulation}
%The subjective experiments from a dedicated test (e.g. for CS 401) led to the following data.
Collected ratings in each of the above-mentioned subjective experiments can be described as following, where $N_{x,u,q}$ reflects the number of ratings on the scale $q \in \{1,\dots,5\}$ given by user $u$ for the degradation condition $x$.
The set of all users is denoted as $\mathcal{U}$ and the MOS~($M_x$) of condition $x$  can be calculated as
\begin{equation}
M_x = \sum_{u \in \mathcal{U}} \frac{\sum_{q=1}^5 q \cdot N_{x,u,q} }{N_{x,u}}
\end{equation}
with $N_{x,u} = \sum_{q=1}^5 N_{x,u,q}$ as the number of ratings from user $u$ for condition $x$.
Thereby, $M_{x,u}=\frac{\sum_{q=1}^5 q \cdot N_{x,u,q} }{N_{x,u}}$ is the MOS value for user $u$ and condition $x$.
It should be noted that in the CS test users may rate the same condition several times (i.e. ratings from different stimuli which refer to the same degradation) or may not rate a certain condition at all.

% CS 701: avg. ratings 0.72 and for non-zeros: 1.40
% CS 501: avg. ratings 1.60 and for non-zeros: 1.93
% CS 401: avg. ratings 3.06 and for non-zeros: 3.53

%Based on the subjective data, simulations were conducted by sampling from the existing data with replacement.
Based on the above-mentioned subjective data, simulations were conducted by randomly sampling with replacement from the collected votes.
%For each test condition $x$, a fixed number $n \in \{10,\dots, 200\}$ of votes were sampled.
For each degradation condition $x$, a fixed number $n \in \{10,\dots, 200\}$ of votes were sampled in two steps.
%In particular, the $n$ users were drawn following the empirical distribution concerning the ratings per user.
First, $n$ users were drawn following the empirical probability $P(U=u|x)$, which, given the condition $x$, estimates a probability that the user $u$ provides a rating for condition $x$:
%The probability $P(U=u|x)$ to sample user $u$ for condition $x$ depends on the number of ratings $N_{x,u}$. % from $u$ for condition $x$.
%depends on the number of ratings $N_{x,u}$. % from $u$ for condition $x$.
\begin{equation}
    P(U=u | x) = \frac{N_{x,u}}{\displaystyle\sum_{u \in \mathcal{U}} N_{x,u}} %\text{, for } u \in \mathcal{U}
\end{equation}
%\babak{I think ", for $u \in \mathcal{U}$" is unnecessary}

%Then, for each simulated user $u$ and condition $x$, the individual rating is sampled from the user rating distribution of user $u$ for condition $x$.
Then, for each selected user $u$ and condition $x$, the individual vote is sampled from the user rating distribution of user $u$ for condition $x$:
\begin{equation}
    P(Q=q|x,u) = \frac{N_{x,u,q}}{\sum_{q=1}^5 N_{x,u,q}} 
\end{equation}

%\babak{@Tobias: what is the benefit of (creating probability distribution functions and) two step sampling compare to pure bootstraping?}
%\tobias{It is needed to sample also which users are rating in order to derive IRR.}
%\babak{fare enough, thanks!}
%Given the accepted data, we simulated cases that $m$ accepted votes per condition are collected, where $m \in [10, n]$, and $n$ is the maximum number of accepted votes collected for that dataset.
%We performed the simulation as following: For each interval, we randomly selected $m$ votes from the pool of accepted votes for each condition (with replacements) and calculated the MOS per condition.  
%Afterwards, we calculated different metrics like Spearman's rank correlation and Root-Mean-Square Error (RMSE) to a \emph{potential best case}. 
%When evaluating \emph{validity} we considered the MOS values collected in laboratory environment to be the potential best case (ground truth) and in case of \emph{reliability} we considered best potential case to be when all votes are used. 
The simulation $m$ was repeated $r=250$ times, and the mean and 95\% confidence interval (CI) of each metric were calculated for further evaluation.

%% file: tabs/tab_datasets.tex
\begin{table*}[t]
% Rafael: \caption{Properties of the datasets that were selected from the ITU-T Rec. P.863 competition pool and used for evaluation.}
% Rafael: proposal
\caption{Datasets selected from the ITU-T Rec. P.863 and used for evaluation (test method was P.800, procedure ACR).}
\label{tab:datasets} 
\begin{tabular*}{1\textwidth}{@{\extracolsep{\fill}}l*{3}{p{4cm}}}
\toprule
                     & \textbf{Dataset 401}                  & \textbf{Dataset 501}             & \textbf{Dataset 701}                            \\ \midrule
\textit{Title}                & Psytechnics P.OLQA test 1& SwissQual P.OLQA SWB 1                        & DOLBY                                  \\ 
%	Owner                & Psytechnics Ltd                                                  & SwissQual License AG  & Dolby Laboratories                     \\ 
%\textit{Date}                 & Jul-08        & Sep-08                                                        & 2013                                   \\ 
%\textit{Test method, procedure}          & P.800, ACR         & P.800, ACR                                                         & P.800, ACR                                  \\ % Rafael: proposed to be deleted if more space is needed 
\textit{Number of conditions} & 48            & 50                                                            & 72                                     \\
\textit{Files per condition}  & 24            & 4                                                             & 16                                     \\ 
\textit{Votes per file, (per condition)}       & 8, (192)             & 24, (96)                                                            & 8, (128)                                      \\ 
\textit{Listeners}            & 32            & 24                                                            & 32                                     \\ 
\textit{Design}               & 6 talkers (3m, 3f)  & 4 talkers (2m, 2f)                                            & 4 talkers (2m, 2f)                     \\ 
\textit{Language}             & British English     & German, Swiss pronunciation                                   & American English                       \\ 
\textit{Number of files}        & 1152                & 200                                                           & 1152                                   \\ 
\textit{Listened through}     & Sennheiser HD25-1   & Grado SR60                                                    & Sennheiser HD 600                      \\ 
\bottomrule
\end{tabular*}
\end{table*}

%% file: 03results.tex
\section{Results}
\label{sec:results}
%\subsection{Overall comparison}
\label{overall_cmp}
The MOS values per condition using all accepted votes obtained from the crowdsourcing tests are compared with the values provided from the corresponding lab based experiments (Tab.~\ref{tab:results:mapped}) using the Root Mean Squared Error~(RMSE) and the Spearman's rank correlation coefficient~(SRCC). 
The results show that there is a high correlation between the MOS values from the crowdsourcing tests and those from the lab tests (Fig.~\ref{fig:result:mapped}). 
Meanwhile it shows that there is a bias and different gradient between crowdsourcing and lab scores.
Applying a first-order mapping significantly decreases the RMSE for CS401 but not for the other tests.

\input{tabs/tab_result_cs_vs_crowd.tex}

\subsection{Validity}\label{validity}
\begin{figure}[htbp] 
  \centering
  \begin{subfigure}[b]{0.5\textwidth}
    \centering
    \includegraphics[width=\textwidth]{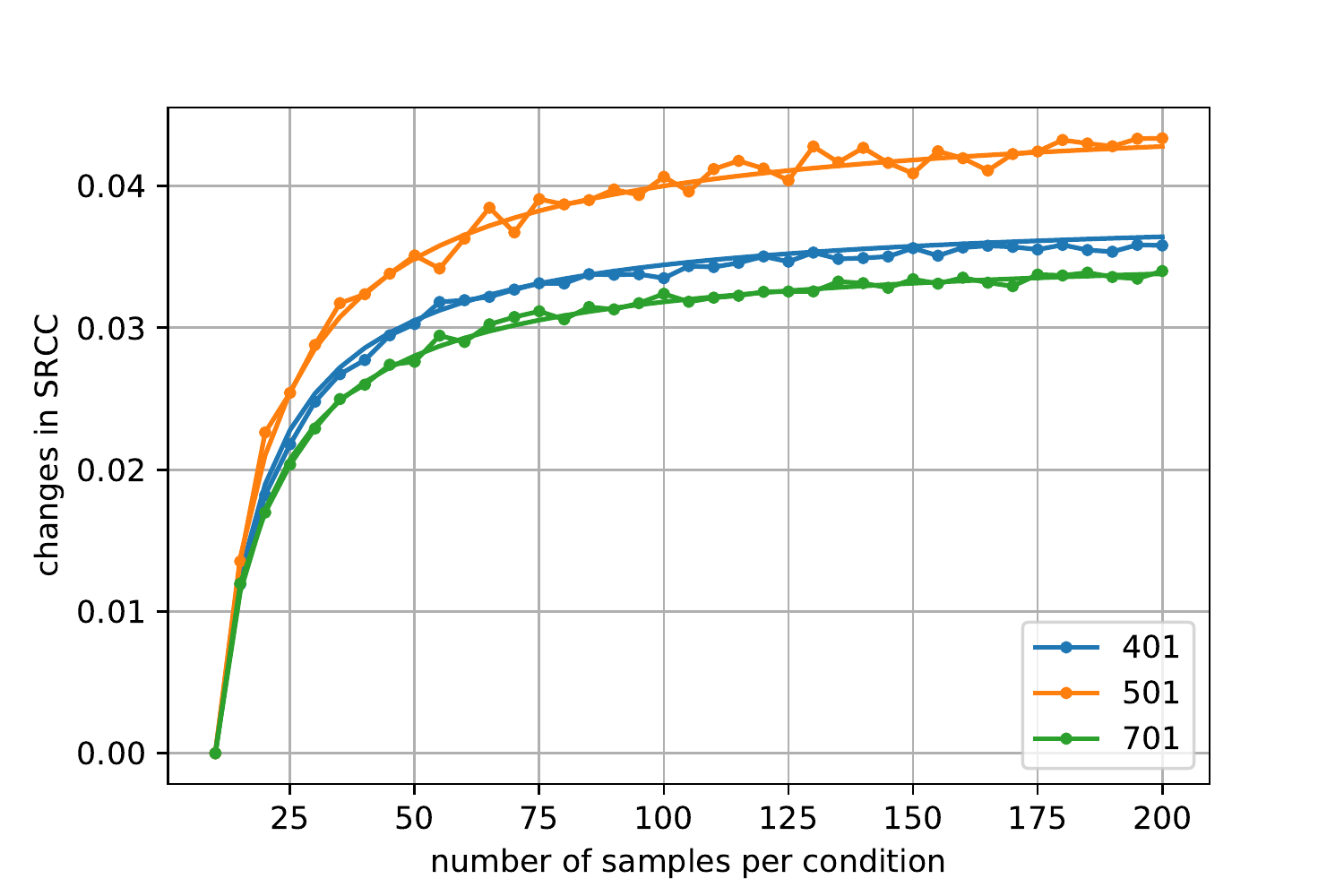} 
    \caption{Changes in Spearman's rank correlation} 
    \label{figV:a} 
  \end{subfigure} 
  ~
  \begin{subfigure}[b]{0.5\textwidth}
    \centering
    \includegraphics[width=\textwidth]{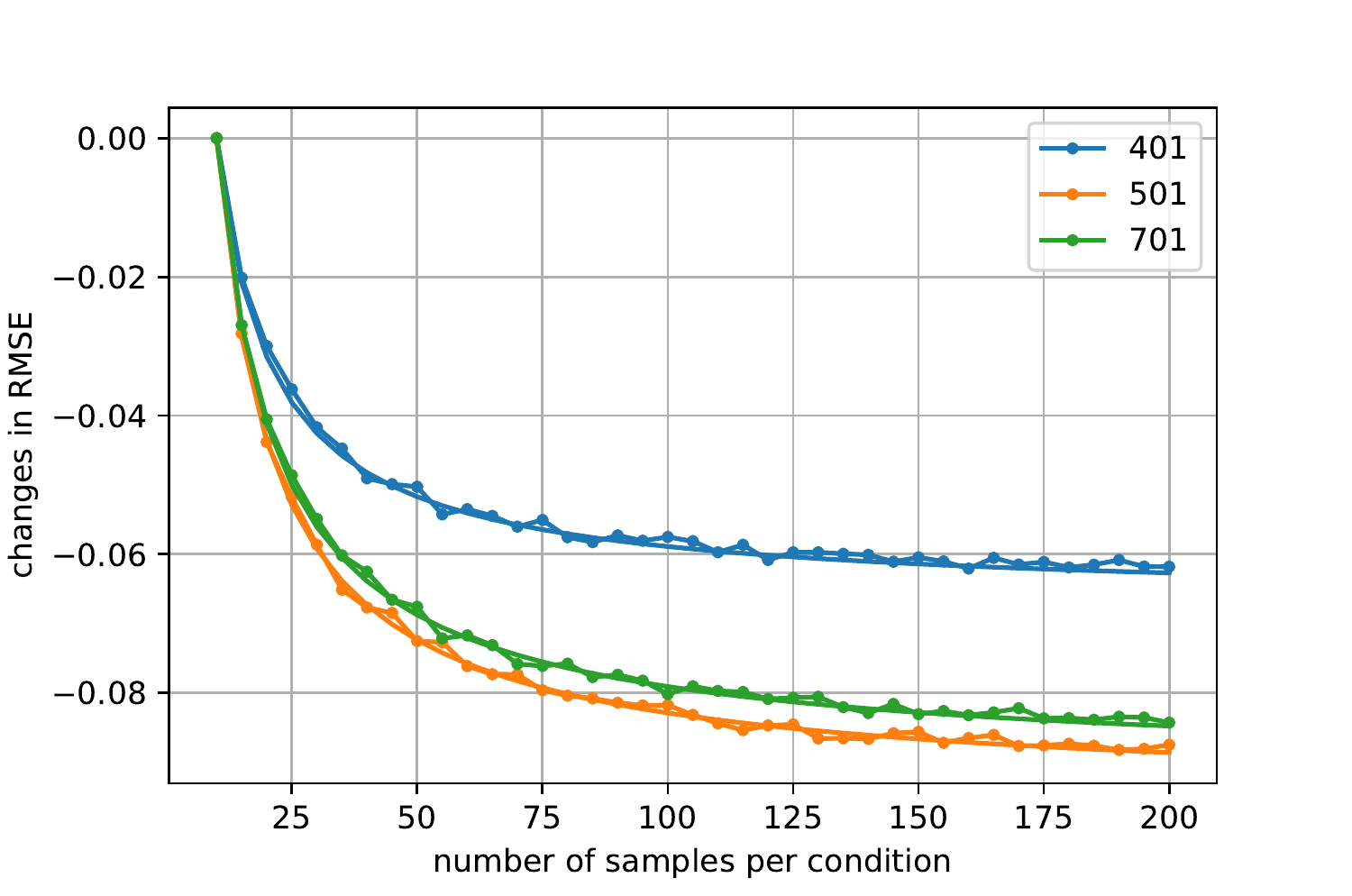}
    \caption{Changes in RMSE}
	\label{figV:b} 
  \end{subfigure}

  \caption{Changes in SRCC (a) and RMSE (b) as a function of number of accepted votes per condition.}
  \label{fig:result:validity} 
\end{figure}

The validity of opinion scores collected through CS as a function of number of votes is examined by calculating the SRCC and RMSE between the calculated MOS values from selected votes in each simulation stage and the MOS values from lab experiments for each dataset. 
Each simulation stage has been repeated 250 times, and the resulting mean and $95\%$ CI for both metrics are used to fit power models for each dataset (Tab.~\ref{tab:results:validity:coef}). 
The changes in SRCC and RMSE as functions of number of votes, as well as the fitted functions are reported in Fig.~\ref{fig:result:validity}. 
%\matthias{Why does the figure not include confidence intervals? If they are so small that they can be neglected, we should state this in the text.}
%The confidence intervals are too small to be visible.
Due to the large number of simulations the average width of the CI is only $0.0007$. Therefore, CI are omitted in the result figures. 
Results show that the curves already flatten out for 60--100 samples per condition.

%\babak{In my simulation with 50 runs, the correlation after 99 votes did not change significantly compared to the correlation with 200 votes. @Tobias: how was for 250? Can you please provide a cutting point?
%}\tobias{See Figure~\ref{fig:result:validity}. The curves already flatten out for 60--100 samples per condition.}

\input{tabs/tab_validity_fit.tex}

\subsection{Certainty Gain}
%The gain in certainty compares the simulated sample with the full CS data. %\matthias{We did not use the abbreviation CS before  -- should be resolved} 
We compared the simulated samples with the full CS data to see how much certainty we gain by adding more ratings.
In other words, we quantify the loss in certainty when only a subset of the ratings is used. 
In the simulation run $i$, we sample $n$ votes for each test condition $x=1,\dots,k$ and calculate the MOS values $m_{x,i}$ for test conditions $x$. %\matthias{I think this notion can be a bit misleading. So far, all MOS values are denoted with a capital M. However, in Section II.C, we define already $M_{x,u}$ as the MOS value for user u and condition x. As the concept of "runs" differs from everything we discussed before about MOS values, I would suggest using a superscription, i.e., $M_{x}^{i}$, to denote the MOS value for condition x obtained in run i }\tobias{That's the reason why it is defined with $m_{x,i}$. We may change the notion later. But since the definition is clearly provided, I don't see a need to change it.}  
The vector $\myvec{m_i} = (m_{x,1}, ... m_{x,k})$ summarizes the MOS values for the sub-sample for all test conditions in the simulation run $i$.% \matthias{Similarly, I suggest to use $M^i$ here}

Meanwhile, MOS value $M_x$, and the corresponding vector $\myvec{M}$, can be calculated using all ratings in the dataset. 
Then, the gain in certainty of simulation run $i$ is quantified as (a) the RMSE and (b) the SRCC between $\myvec{m_i}$ and $\myvec{M}$. Again, the simulation is repeated $r=250$ times 
%\florian{this statement has been repeated multiple times; remove some?} \tobias{No, I want to stress that the average is computed over the simulation runs}
and the average gain in certainty is derived over the simulation runs. 
\begin{align}
    G(n) = \frac{1}{r} \sum_{i=1}^r G_i(n),  && G_i(n) & =  SRCC(\myvec{m_i},\myvec{M})\\
    G^*(n) = \frac{1}{r} \sum_{i=1}^r G_i^*(n), && G_i^*(n) & =  RMSE(\myvec{m_i},\myvec{M})
\end{align}
%\babak{is it not more readable to put  $G(n)$ in left and $G_i(n)$ on right side? }
%\tobias{That's fine for me.}

\begin{figure}[htbp] 
  \centering
  \begin{subfigure}[b]{0.5\textwidth}
    \centering
    \includegraphics[width=\textwidth]{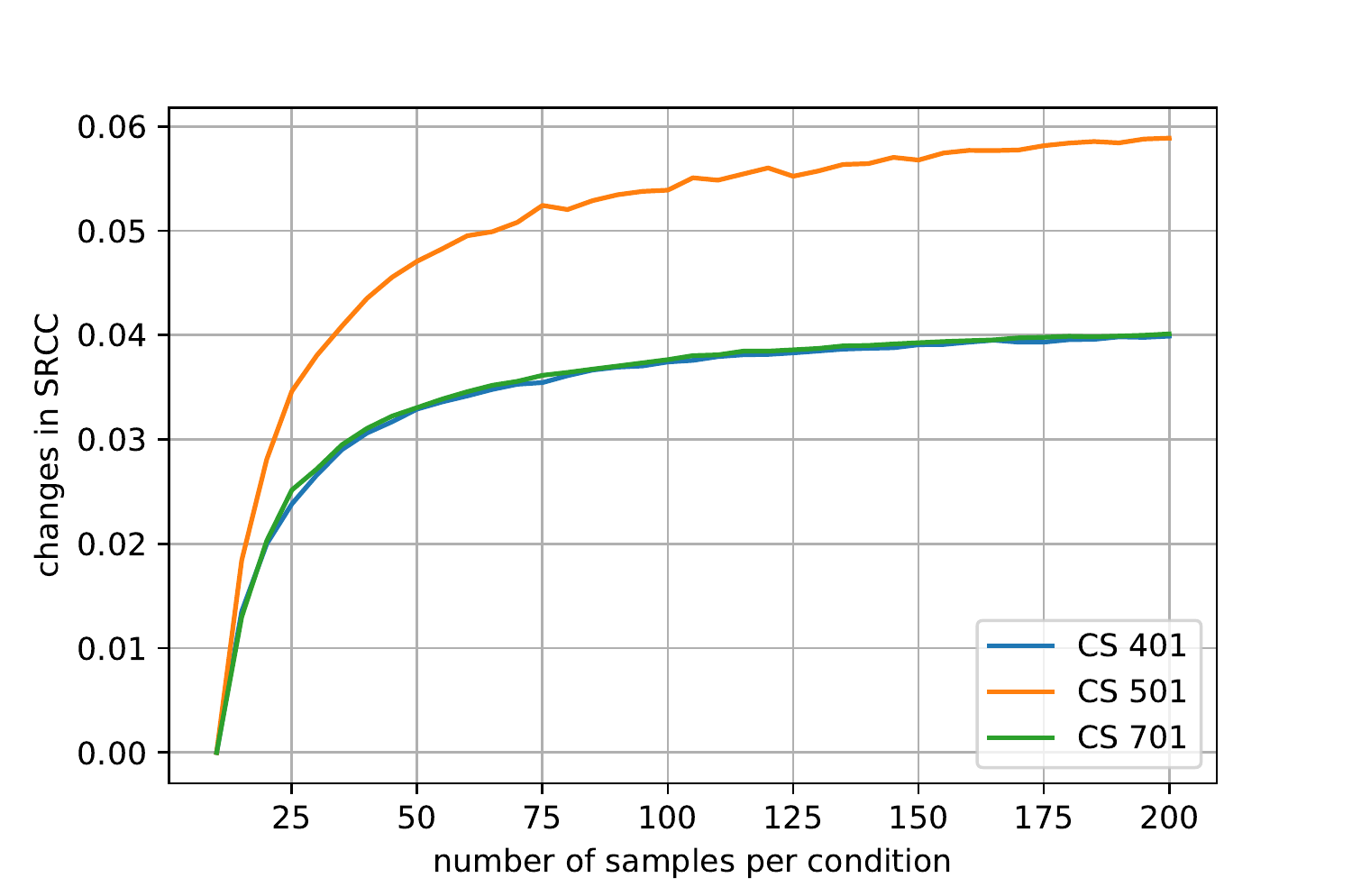} 
    \caption{Changes  $\Delta G(n)$ in Spearman's rank correlation} 
    \label{fig:certainty:srcc} 
  \end{subfigure} 
  ~
  \begin{subfigure}[b]{0.5\textwidth}
    \centering
    \includegraphics[width=\textwidth]{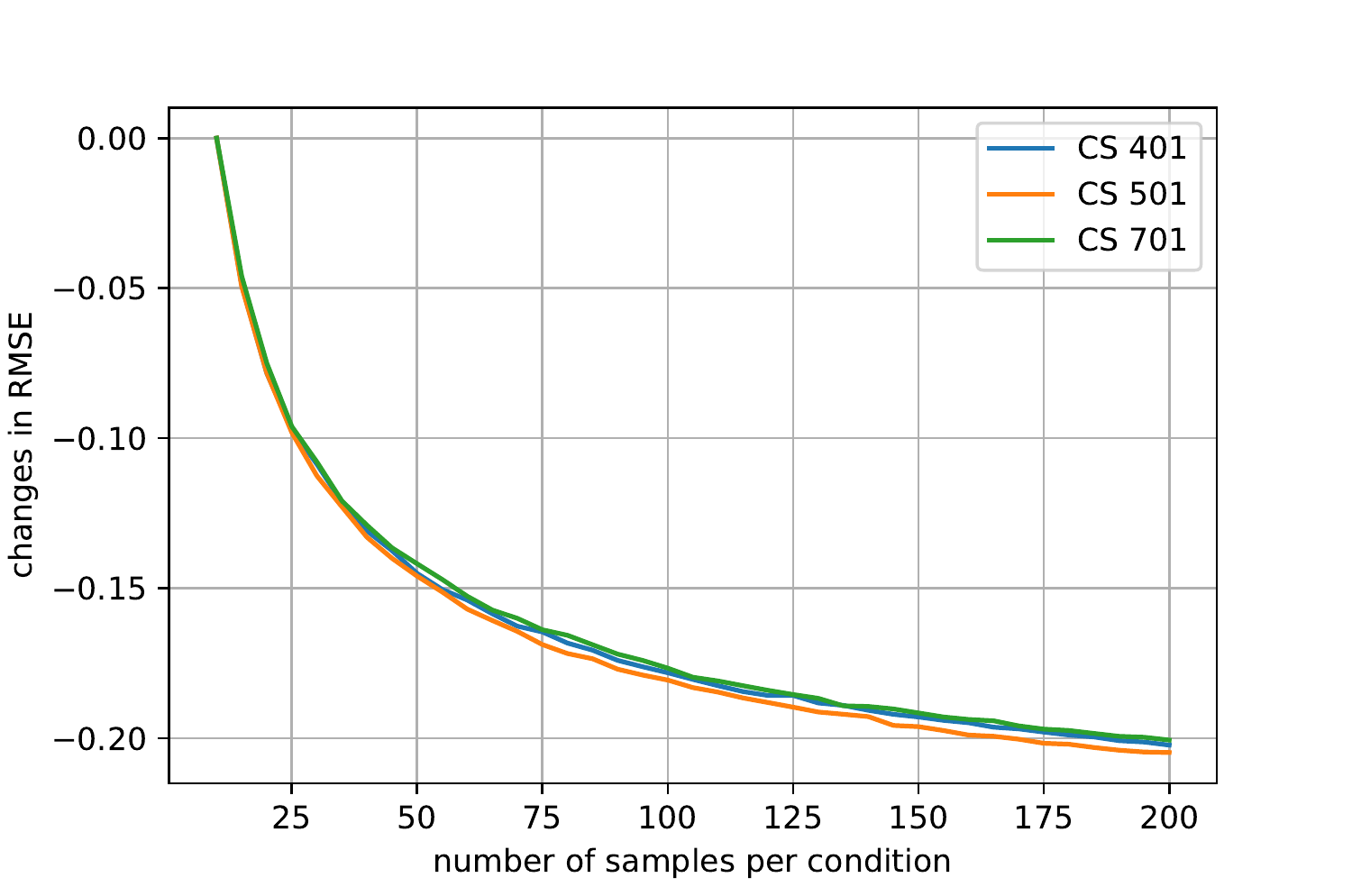}
    \caption{Changes $\Delta G^*(n)$ in RMSE}
	\label{fig:certainty:rmse} 
  \end{subfigure}

  \caption{Certainty gain: Changes $\Delta G(n)$ in SRCC (a) and changes $\Delta G^*(n)$ in RMSE (b) as a function of number of accepted votes per condition.}
  \label{fig:certainty} 
\end{figure}

Fig.~\ref{fig:certainty} shows the changes in the certainty gain for SRCC and RMSE. %\matthias{Would CIs in the figure be meaningful here?} 
In particular, the changes are computed by subtracting the values by the first point for a sample size of $n=10$, i.e., $\Delta G(n) = G(n)-G(10)$ and $\Delta G^*(n) = G^*(n)-G^*(10)$. %\florian{I really do not understand this sentence. What is shifted here, and why and how (and why "in particular")? Can you rephrase this?}
%\babak{it is to show the "gain", considering 10 votes be the base, how much do we gain if we go for e.g.50 votes.}
First, we observe that the changes in the SRCC are almost similar for CS 401 and CS 701. For CS 501, the obtained changes in the SRCC are higher. 
%\tobias{why?}. 
For more than 50 samples per condition, the curves start to flatten out, thus, around 60--100 samples seem to provide a good recommendation. For the changes in the RMSE, there are no significant curves across the datasets. 
When $n>60$, the difference between resulting RMSE and the RMSE with $n=10$ votes is more than 0.15~MOS.
%\babak{do not get the last sentence.}\tobias{Clear now?}

%\todo{describe figure}

%\todo{The correlation after 99 votes did not change significantly compared to the correlation with 200 votes.
%}

\subsection{Confidence Interval Width}
Another measure for the uncertainty of the data considers the width of the CI for the MOS values. In particular, in the simulation run $i$, we sampled $n$ votes for each condition $x=1,\dots,k$. The width $W_{x,i}(n)$ of the CI for the MOS of condition $x$ in run $i$ is computed based on bootstrapping, as recommended in \cite{hossfeld2018confidence}, and is the difference between the upper and the lower CI bound. 
We used the non-parametric bootstrap method by Efron \cite{efron1992bootstrap}, which solely employs the empirical distribution of the user ratings for a condition $x$ to derive the (not necessarily symmetric) CI around the mean. Then, we obtain the following average CI width as uncertainty measure.
\begin{align}
    W(n) = \frac{1}{r} \sum_{i=1}^r W_i(n), && W_i(n) & = \sum_{x=1}^{k}  W_{x,i}  
\end{align}

\begin{figure}
    \centering
    \includegraphics[width=\columnwidth]{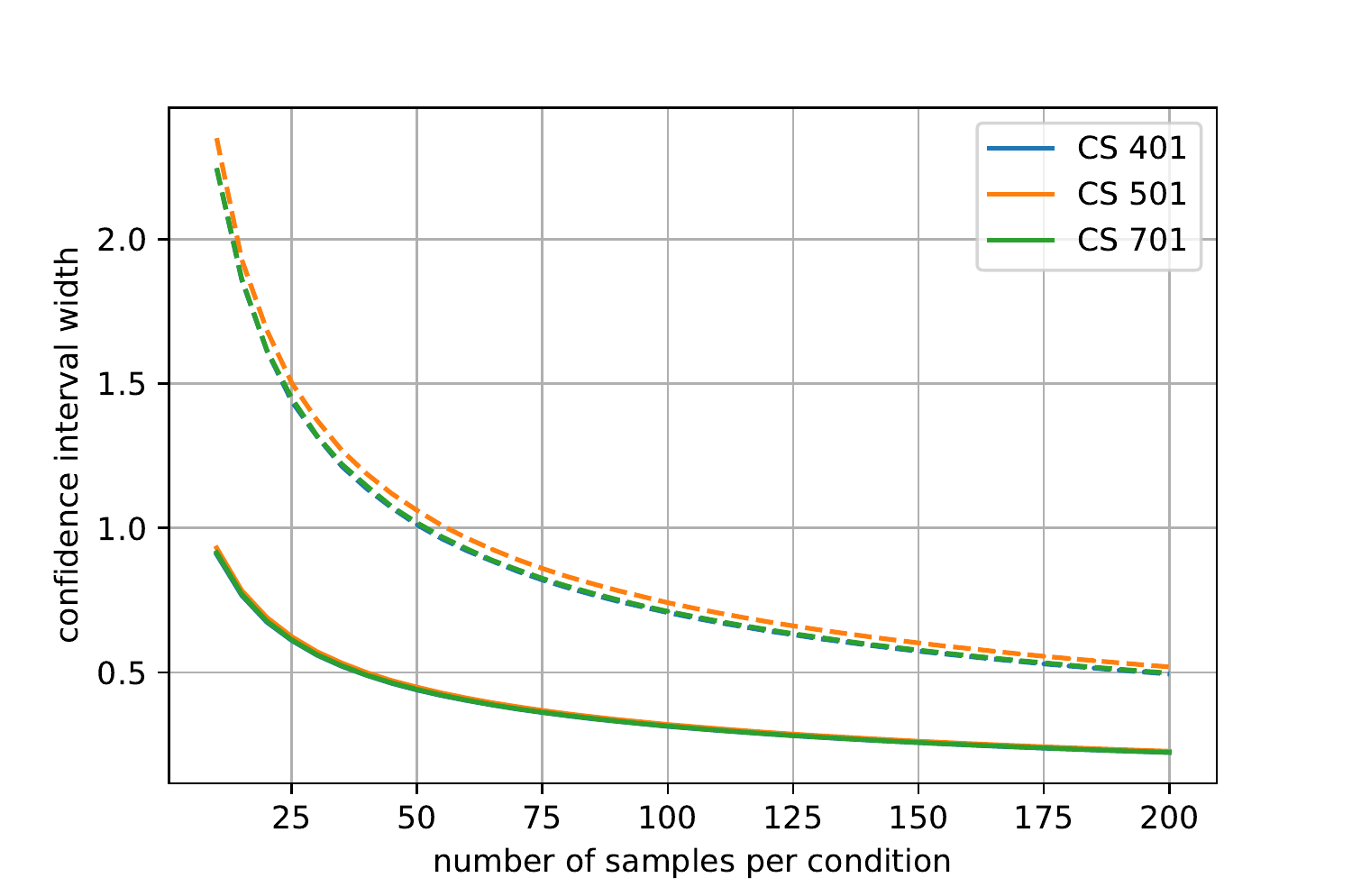}
    \caption{Confidence interval width $W$ as measure for certainty of the CS data sets (solid line) as well as the maximum possible value $W$ assuming maximum variance (dashed line). %\tobias{Maximum CI added in the lower figure (dashed line). The maximum conf. interval is derived by assuming that a ratio of $p$ users rates 5 and $1-p$ rate 1. For a given MOS $\mu$, it is $p=\frac{\mu-1}{4}$. Then, binomial proportion confidence intervals are computed for a given sample size (via Clopper-Pearson intervals \cite{hossfeld2018confidence} and we obtain $p_L,p_H$. Then the MOS confidencence interval width is $(p_H\cdot 5+(1-p_H)\cdot 1) - (p_L\cdot 5+(1-p_L)\cdot 1) = 4 \cdot (p_H-p_L )$.}
    }
    \label{fig:confidence}
\end{figure}

Fig.~\ref{fig:confidence} plots the confidence interval width depending on the number of samples per condition $n$ for the three different datasets. We can observe that $W(n)<0.4$ for $n>60$. For $W(n)<0.3$, at least $n=115$ samples per condition are required. %\babak{can you please report $W(n)<0.3$? Based on ITU-T P.1401 it is typical width for subjective tests.} 
It can be seen that the results are almost identical across the datasets. In the lab experiments, the average confidence interval width is 0.20, 0.31, 0.26 for the lab dataset 401, 501, 701 based on 192, 96, 128 users, respectively.
%\tobias{conf. int. width of 0.4 seems to be okay.} \matthias{Tobias, you always had this nice example that, due to the bounded MOS scale, one can create arbitrarily small CI intervals using CS. Therefore, we always argued that additional reliability checks need to be implemented. I think in this context here, it would be interesting to also calculate the "worst case" CI length (for our number of participants), to illustrate how good our results are.}\tobias{Good idea: I added Figure~\ref{fig:confidence}.}
%babak: well done!

Fig.~\ref{fig:confidence} plots also the maximum possible CI width as dashed line. This maximum CI is obtained, when a ratio $p$ of users rates $5$ and the ratio $1-p$ of the remaining users  rate $1$, see \cite{hossfeld2011sos}. 
%\babak{last sentence looks incomplete.} 
For a given MOS value $\mu$, it is $p=\frac{\mu-1}{4}$. 
Then, binomial proportion CIs are computed for a given sample size based on Clopper-Pearson intervals \cite{hossfeld2018confidence}, resulting in a lower bound $p_L$ and an upper bound $p_H$, respectively. Then, the MOS confidence interval width is $4 \cdot (p_H-p_L )$. %(p_H\cdot 5+(1-p_H)\cdot 1) - (p_L\cdot 5+(1-p_L)\cdot 1) = 4 \cdot (p_H-p_L )$.
%\babak{is the formula needed?}
%\tobias{No, not needed, but may be nice to follow. Maybe shorten to $4 \cdot (p_H-p_L )$.}

\newcommand{\irr}{\Gamma}
\subsection{Inter-Rater Reliability (IRR)}
The inter-rater reliability~(IRR) $\irr_i(n)$ %\matthias{Previously, we used $r$ for the number of simulation runs. Maybe, we should not use a capital R and a small r for unrelated metrics. } 
is derived for a single simulation run $i$ where each condition is rated $n$ times. An individual user $u$ rates several conditions $\mathcal{X}_u$ with corresponding MOS value $M_{x,u}$ for \mbox{$x \in \mathcal{X}_u$}. For the conditions $x$, the MOS value $M_{x,\mathcal{U}\setminus{u}}$ %$M_{\mathcal{U}\setminus{u},x}$ 
%\matthias{I think it should be: $M_{x,\mathcal{U}\setminus{u}}$} 
is derived by considering all other users except $u$. Then, the SRCC is computed for all conditions $x \in \mathcal{X}_u$ between the MOS $M_{x,u}$ from user $u$ and the MOS from all others $M_{x,\mathcal{U}\setminus{u}}$. %$M_{\mathcal{U}\setminus{u},x}$  \matthias{I think it should be: $M_{x,\mathcal{U}\setminus{u}}$}. 
The inter-rater reliability $\irr(n)$ of simulation run $i$ considers then SRCC values averaged over all users.
%\babak{what if the user did not rate condition k? corr = 0? }\tobias{Then, this condition is not considered for the user to compute the IRR value for that user. We compute SRCC of MOS of user $u$ for the seen test conditions of that user. This is compared with the MOS from the other users.}
\begin{equation}
    \irr_i(n) = \frac{1}{\left|\mathcal{U}\right|}\sum_{u \in \mathcal{U}}SRCC(M_{\mathcal{X}_u,u}, M_{x,\mathcal{U}\setminus{u}})
\end{equation}
%\matthias{I think it should be: $M_{x,\mathcal{U}\setminus{u}}$}

\begin{figure}
    \centering
    \includegraphics[width=\columnwidth]{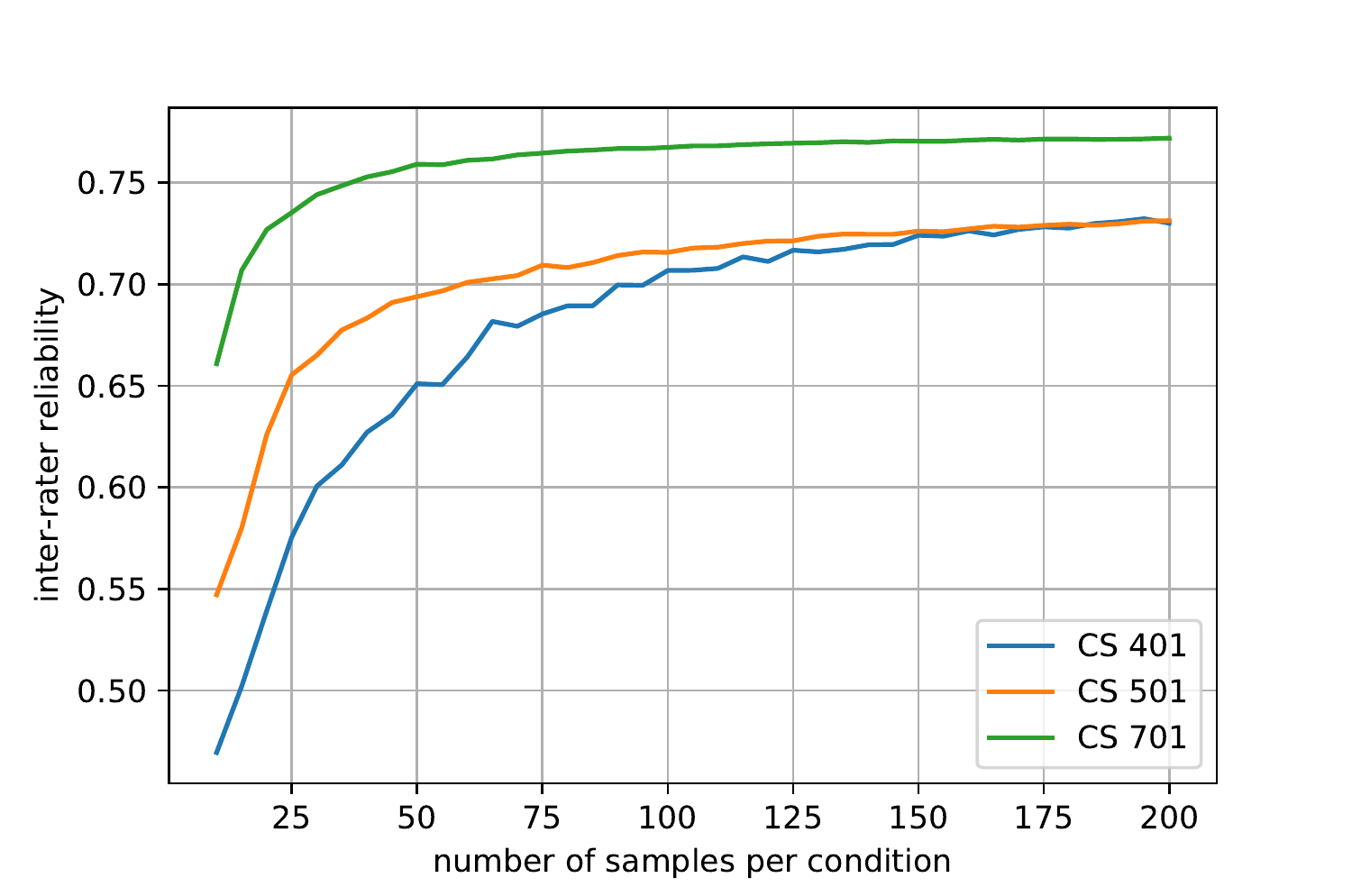}
    \caption{Inter-rater reliability $\irr$ for the crowdsourcing data sets.}% For the different datasets, the IRR for the entire data is as follows. }
    \label{fig:irr}
\end{figure}

The inter-rater reliability $\irr(n)$ is again the average over $r=250$ repetitions, $\irr(n) = \frac{1}{r}\sum_{i=1}^r \irr_i(n)$. Fig.~\ref{fig:irr} depicts the IRR depending on the sample size $n$.
We see significant differences between all three curves in terms of absolute IRR. However, the shape of the curves is again similar. After 60--100 the curves flatten out once more. Note that the IRR is bounded by the IRR of the entire CS dataset ($\irr_{CS 401}=0.7945, \irr_{CS 501}=0.7453, \irr_{CS 701}= 0.7773$). %\matthias{Should we include these numbers here?}. 
\babak{We can also clearly see that below 40 samples, the IRR significantly drops.} We recommend to have at least 60 samples with regard to the IRR.

% %\subsection{Reliability}\label{reliabilty}
% %\subsubsection{no-name}
% %\begin{itemize}
% %  \item Same simulation: x votes are randomly selected where x∈[10,n], and MOS values are calculated.
% %
% %  \item The charts shows the correlation and RMSE of MOS values with x votes with MOS values with the case with n (all) votes.
% %
% %\end{itemize}

% \begin{figure}[htbp] 
%   \centering
%   \begin{subfigure}[b]{0.5\textwidth}
%     \centering
%     \includegraphics[width=\textwidth]{figs/irr_r.png} 
%     \caption{Changes in Spearman's rank correlation} 
%     \label{fig:r:a} 
%   \end{subfigure} 
%   ~
%   \begin{subfigure}[b]{0.5\textwidth}
%     \centering
%     \includegraphics[width=\textwidth]{figs/irr_rmse.png}
%     \caption{Changes in RMSE}
% 	\label{fig:r:b} 
%   \end{subfigure} 

%   \caption{Changes in Spearman's rank correlation (a) and RMSE (b) as a function of number of accepted votes per condition comparing XXXX. \todo{figures need to have a same style}}
%   \label{fig:result:reliability} 
% \end{figure}

%% file: tabs/tab_result_cs_vs_crowd.tex
\begin{table}[b]
\caption{Comparison between MOS values obtained through Crowdsourcing and Laboratory (total accepted votes).}
\label{tab:results:mapped} 
\begin{center}
\begin{tabular}{ccccc}
\toprule
{\multirow{2}*{\textbf{Test}}} & {\multirow{2}*{\textbf{SRCC}}} & {\multirow{2}*{\textbf{RMSE}}} & {\multirow{2}*{\textbf{RMSE -FOM\textsuperscript{b}}}} & \textbf{Avg. votes}\\
                            &   & & & \textbf{p. cond. (STD)}\\

\midrule
CS 401 & $.971$\textsuperscript{a} & $0.485$ & $0.169$  & $217$ ($4.8$) \\ 
CS 501 & $.891$\textsuperscript{a} & $0.324$ & $0.316$  & $102$ ($7.3$) \\ 
CS 701 & $.931$\textsuperscript{a} & $0.32$ & $0.318$  & $113$ ($8.8$) \\ \bottomrule
\multicolumn{3}{c}{$^{\mathrm{a}\ } p < .001$} & \multicolumn{2}{l}{\textsuperscript{b} after 1\textsuperscript{st} order mapping}
\end{tabular}
\end{center}
\end{table}

%% file: tabs/tab_validity_fit.tex
\begin{table}[htbp]
\caption{Coefficient of Power models used ($y=a\cdot x^b+c$), where $x$ is the number of votes and $y$ is the predicted value. The model parameters ($a_1, b_1, c_1$) are also provided when using first order mappings of the crowdsourced MOS results.}
\label{tab:results:validity:coef} 
\begin{center}
\begin{tabular}{ccccccc}
\toprule
\textbf{Model} & \textbf{a} & \textbf{b} & \textbf{c} & $\mathbf{a_1}$ & $\mathbf{b_1}$ & $\mathbf{c_1}$\\
\midrule
%$r_s 401$ & -0.3707 & -0.9912 & 0.9752 \\ \hline
%$r_s 501$ & -0.3901 & -0.913 & 0.891 \\ \hline
%$r_s 701$ & -0.3857 & -0.9937 & 0.9272 \\ \hline
SRCC 401 & -0.3837 &  -1.0129 &  0.9749 & -0.3837 &  -1.0129 &  0.9749 \\
SRCC 501 & -0.3039 &  -0.8319 &  0.8916 & -0.3039 &  -0.8319 &  0.8916 \\
SRCC 701 & -0.3443 &  -0.9675 &  0.9317 & -0.3443 &  -0.9675 &  0.9317 \\
\midrule
%RMSE 401 & 0.7687 & -0.8342 & 0.1668 \\ \hline
%RMSE 501 & 0.5354 & -0.796 & 0.3119 \\ \hline
%RMSE 701 & 0.777 & -0.8598 & 0.3144 \\ \hline
RMSE 401 & 0.6467 &  -0.9903 &  0.4803 & 0.8004 &  -0.8306 &  0.1647 \\
RMSE 501 & 0.6717 &  -0.8544 &  0.3184 & 0.6528 &  -0.8588 &  0.3109 \\
RMSE 701 & 0.7667 &  -0.9142 &  0.3172 & 0.8141 &  -0.9074 &  0.3149 \\
\bottomrule
\end{tabular}
\end{center}
\end{table}
%\babak{It refers to pure MOS not mapped one. In case we want to report 1st ordered mapped one, just the coefficient $c$ should be updated.}
%\tobias{I added the power model coefficients also when using first order mappings.}

%% file: 04discussion.tex
\section{Discussion and Conclusion}
\label{sec:discussion}

We performed an analysis of quality judgments obtained in three crowdsourcing studies carried out on three different standard databases, using different crowdsourcing platforms and experimenters. 
For each database, corresponding lab judgments were provided which were collected according to ITU-T Rec. P.800. 
We consider the lab data as a ground truth for our crowdsourcing experiments, acknowledging that lab data also carries an uncertainty, and that the lab environment may not lead to ecologically valid results.

Regarding the validity of the crowdsourcing experiments, in all cases a good correlation to the lab data could be achieved.
Spearman's rank order correlation ranged between $0.89$ and $0.97$ for the three databases, and the RMSE between $0.48$ and $0.32$ on the MOS scale. 
For the 401 database, a bias between lab and CS data was observed, which could be removed with a 1st order mapping, resulting in a rather low RMSE ($0.17$). 
Overall, taking the lab data as a ``gold standard'', a good validity of the crowdsourcing experiments could be achieved.

When varying the number of votes (which was the focus of the experiments), both correlation and RMSE did not substantially increase after approx. $60$ votes per condition. 
Although the absolute value of both metrics and the relative increase of the metric with increasing number of votes below 60 votes is different for each database (especially for the 701 database), the increase in validity does not seem to justify the cost of having more than 60 votes per condition.

Regarding the reliability, a similar picture arises.  
Both the confidence interval width and the inter-rater reliability do not substantially change with more than $60$ ratings per condition. 
Again, database 701 sticks out in that it has a higher inter-rater reliability score than the other two databases.

Regarding the observed differences between the databases, there are a number of factors which could be their cause. 
%dataset or subjective ratings collected for datasets?
Database 501, for which the CS results showed the lowest correlation to the lab, contained Swiss-German samples, which were rated in Switzerland in the lab, whereas in the crowd most crowdworkers were from Germany ($90.6\%$). 
%\babak{@rafael: please confirm} \rafael{DONE!}
Small language differences as well as different quality backgrounds (cf. the discussion about references in the introduction) might have caused the lower correlation. Database 701 showed the highest inter-rater reliability amongst all crowdsourcing experiments. 
When carrying out the respective experiment, a rather large proportion of workers was removed from the data, either because of self-reported hearing loss or because these workers answered incorrectly to an attention and device checking question. 
This rather strong cleansing of the data might have lead to the higher inter-rater reliability on the remaining data.

Overall, recommending to use at least 60 votes per test condition in a crowdsourcing speech quality assessment experiment seems to be a reasonable compromise between effort spent and both reliability and validity of the results. \babak{to remove:This value might be considered for inclusion in the ITU-T Rec. P.808.}
\matthias{Same comment as in the introduction: Is it meaningful to include this explicit  purpose here? The paper itself has a scientific value, maybe also for other cs studies outside the standard? }

\babak{Potential aspects to be addressed either here or in the journal paper:
%\begin{itemize}
  %\item 
  generalizability of simulation result? We considered that the ratings we have is a "population" and performed n sampling. 
    %\item 
  Can we extend the current suggestion to file level as well? There are many case that one might be just interested on ratings per file (rather condition).
  %\item Why Inter-rater reliability of 701 is so higher? %\babak{@Matthias: did you use outlier detection methods or compare the distribution of ratings in your data screening process?}
  %  \matthias{No}
  %\item 
  why in figure \ref{fig:result:validity} the changes in RMSE of 401 is so diiferent that the others? %\babak{@Tobias, did you use}
   %\end{itemize}
   }
  \tobias{This is because I was using the true MOS values, not the first order mapped values.}
  \babak{should not be so much matter, as it refers to the difference. I checked my plots, same relationship is there with 1st order mapped MOS.}

%% file: 00main.bbl
\begin{thebibliography}{10}
\providecommand{\url}[1]{#1}
\csname url@samestyle\endcsname
\providecommand{\newblock}{\relax}
\providecommand{\bibinfo}[2]{#2}
\providecommand{\BIBentrySTDinterwordspacing}{\spaceskip=0pt\relax}
\providecommand{\BIBentryALTinterwordstretchfactor}{4}
\providecommand{\BIBentryALTinterwordspacing}{\spaceskip=\fontdimen2\font plus
\BIBentryALTinterwordstretchfactor\fontdimen3\font minus
  \fontdimen4\font\relax}
\providecommand{\BIBforeignlanguage}[2]{{%
\expandafter\ifx\csname l@#1\endcsname\relax
\typeout{** WARNING: IEEEtran.bst: No hyphenation pattern has been}%
\typeout{** loaded for the language `#1'. Using the pattern for}%
\typeout{** the default language instead.}%
\else
\language=\csname l@#1\endcsname
\fi
#2}}
\providecommand{\BIBdecl}{\relax}
\BIBdecl

\bibitem{Jekosch05}
U.~Jekosch, \emph{Voice and Speech Quality Perception. Assessment and
  Evaluation}.\hskip 1em plus 0.5em minus 0.4em\relax Berlin: Springer, 2005.

\bibitem{Lienert89}
G.~A. Lienert, \emph{Testaufbau und {T}estanalyse}.\hskip 1em plus 0.5em minus
  0.4em\relax Weinheim: Verlag Julius Beltz, 1989.

\bibitem{ITU-P800}
{ITU-T Recommandation P.800}, \emph{{Methods for Subjective Determination of
  Transmission Quality}}.\hskip 1em plus 0.5em minus 0.4em\relax Geneva:
  International Telecommunication Union, 1996.

\bibitem{naderi2015effect}
B.~Naderi, T.~Polzehl, I.~Wechsung, F.~K{\"o}ster, and S.~M{\"o}ller, ``{Effect
  of Trapping Questions on the Reliability of Speech Quality Judgments in a
  Crowdsourcing Paradigm},'' in \emph{Conference of the International Speech
  Communication Association}, 2015.

\bibitem{polzehl2015robustness}
T.~Polzehl, B.~Naderi, F.~K{\"o}ster, and S.~M{\"o}ller, ``Robustness in speech
  quality assessment and temporal training expiry in mobile crowdsourcing
  environments,'' in \emph{Conference of the International Speech Communication
  Association}, 2015.

\bibitem{Egger-Lampl2017a}
S.~Egger-Lampl, J.~Redi, T.~Ho{\ss}feld, M.~Hirth, S.~M{\"{o}}ller, B.~Naderi,
  C.~Keimel, and D.~Saupe, ``{Crowdsourcing Quality of Experience
  Experiments},'' in \emph{Evaluation in the Crowd. Crowdsourcing and
  Human-Centered Experiments}, D.~Archambault, H.~Purchase, and T.~Ho{\ss}feld,
  Eds.\hskip 1em plus 0.5em minus 0.4em\relax Cham: Springer International
  Publishing, 2017.

\bibitem{ZequeiraJimenez2018}
R.~{Zequeira Jim{\'{e}}nez}, L.~{Fern{\'{a}}ndez Gallardo}, and
  S.~M{\"{o}}ller, ``{Influence of Number of Stimuli for Subjective Speech
  Quality Assessment in Crowdsourcing},'' in \emph{International Conference on
  Quality of Multimedia Experience}, 2018.

\bibitem{ITU-P808}
{ITU-T Recommandation P.808}, \emph{{Subjective Evaluation of Speech Quality
  with a Crowdsourcing Approach}}.\hskip 1em plus 0.5em minus 0.4em\relax
  Geneva: International Telecommunication Union, 2018.

\bibitem{ITU-P863}
{ITU-T Recommandation P.863}, \emph{{Perceptual Objective Listening Quality
  Prediction}}.\hskip 1em plus 0.5em minus 0.4em\relax Geneva: International
  Telecommunication Union, 2018.

\bibitem{smits2004development}
C.~Smits, T.~S. Kapteyn, and T.~Houtgast, ``Development and validation of an
  automatic speech-in-noise screening test by telephone,'' \emph{International
  journal of audiology}, vol.~43, no.~1, pp. 15--28, 2004.

\bibitem{hossfeld2018confidence}
T.~Hossfeld, P.~E. Heegaard, M.~Varela, and L.~Skorin-Kapov, ``{Confidence
  Interval Estimators for MOS Values},'' \emph{arXiv preprint
  arXiv:1806.01126}, 2018.

\bibitem{efron1992bootstrap}
B.~Efron, ``{Bootstrap Methods: Another Look at the Jackknife},'' in
  \emph{Breakthroughs in Statistics}.\hskip 1em plus 0.5em minus 0.4em\relax
  Springer, 1992.

\bibitem{hossfeld2011sos}
T.~Ho\ss{}feld, R.~Schatz, and S.~Egger, ``{SOS: The MOS is not Enough!}'' in
  \emph{International Workshop on Quality of Multimedia Experience}, 2011.

\end{thebibliography}
